\documentclass[11pt]{article}
\usepackage{graphicx,amsmath,amsfonts,amssymb}
\usepackage{color,multicol,multirow}
\usepackage{array}
\def\sloppy{\tolerance=100000\hfuzz=\maxdimen\vfuzz=\maxdimen}
\def \beq  {\begin{equation}}
\def \eeq  {\end{equation}}
\def \beqar {\begin{eqnarray}}
\def \eeqar {\end{eqnarray}}
\def\bsp{\beq\begin{split}}
\allowdisplaybreaks
\mathchardef\mhyphen="2D

\def\del {\partial}

\def \ch {{\rm ch}}
\def \sh {{\rm sh}}
\def \th {{\rm th}}

\def\O {{\cal O}}

\def\half{\frac{1}{2}}

\begin{document}
\def \CMP {{Commun. Math. Phys.}}
\def \PRL {{Phys. Rev. Lett.}}
\def \PL {{Phys. Lett.}}
\def \NPBProc {{Nucl. Phys. B (Proc. Suppl.)}}
\def \NP {{Nucl. Phys.}}
\def \RMP {{Rev. Mod. Phys.}}
\def \JGP {{J. Geom. Phys.}}
\def \CQG {{Class. Quant. Grav.}}
\def \MPL {{Mod. Phys. Lett.}}
\def \IJMP {{ Int. J. Mod. Phys.}}
\def \JHEP {{JHEP}}
\def \PR {{Phys. Rev.}}
\begin{titlepage}
\null\vspace{-62pt} \pagestyle{empty}
\begin{center}
\rightline{CCNY-HEP-14/1}
\rightline{March 2014}
\vspace{1truein} {\Large\bfseries
Exact operator Hamiltonians and interactions in the droplet bosonization method}\\
\vskip .1in
{\Large\bfseries ~}\\
{\large\sc Dimitra Karabali$^a$} and
 {\large\sc Alexios P. Polychronakos$^b$}\\
\vskip .2in
{\itshape $^a$Department of Physics and Astronomy\\
Lehman College of the CUNY\\
Bronx, NY 10468}\\
\vskip .1in
{\itshape $^b$Physics Department\\
City College of the CUNY\\
New York, NY 10031}\\
\vskip .1in
\begin{tabular}{r l}
E-mail:&{\fontfamily{cmtt}\fontsize{11pt}{15pt}\selectfont dimitra.karabali@lehman.cuny.edu}\\
&{\fontfamily{cmtt}\fontsize{11pt}{15pt}\selectfont alexios@sci.ccny.cuny.edu}
\end{tabular}

\fontfamily{cmr}\fontsize{11pt}{15pt}\selectfont
\vspace{.8in}
\centerline{\large\bf Abstract}
\end{center}
We derive the exact form of the bosonized Hamiltonian for a many-body fermion system in one
spatial dimension with
arbitrary dispersion relations, using the droplet bosonization method. For a single-particle Hamiltonian
polynomial in the momentum, the bosonized Hamiltonian is a polynomial of one degree higher
in the bosonic `boundary' field and includes subleading lower-order and derivative terms.
This generalizes the known results for massless relativistic and nonrelativistic fermions (quadratic
and cubic bosonic Hamiltonians, respectively). We also consider two-body interactions and demonstrate
that they lead to interesting collective behavior and phase transitions in the Fermi sea.

\end{titlepage}

\pagestyle{plain} \setcounter{page}{2}
\setcounter{footnote}{0}
\setcounter{figure}{0}
\renewcommand\thefootnote{\mbox{\arabic{footnote}}}

\section{Introduction}

Bosonization, the description of fermion systems in terms of bosonic degrees
of freedom, is a well-established and useful technique, at least in one space dimension
\cite{Early,Bozo,WZW}.
Such a description generically maps many-body collective excitations of the fermion system
into large coherent bosonic excitations. Some quantum features of the fermion system
are therefore mapped to classical features of the bosonic system in a way analogous to
duality transformations. Because of this, bosonization is a convenient setting to study strongly
correlated systems of fermions.

A particularly intuitive approach to bosonization is a hydrodynamic phase space
formulation of the many fermion system, otherwise known as the ``droplet" method.
In this approach we start with a semiclassical description of the many-body system
in terms of a dense collection of particles in their single-particle phase space, forming
a constant-density distribution inside a (generically connected) domain, the `droplet'.
The boundary of this domain essentially corresponds to the Fermi surface,
and the dynamics of this boundary encodes the many-body degrees of freedom \cite{droplet,phase}.
In one dimension, this
leads to a chiral theory that fully reproduces the standard bosonization
results. In particular, for nonrelativistic fermions, it reproduces the
results of the collective field method \cite{collect}. Fermion excitations 
around the Fermi (or Dirac) sea become one-dimensional waves with the
corresponding phonon
states encoding quantum excitations, while properly defined exponentials in the
bosonic field (`vertex operators') become fermion operators. Overall, we have a perturbatively
exact mapping of states between the two systems. For systems with a finite number
of fermions this mapping can fail at the nonperturbative level, when the Fermi sea
is completely depleted. For relativistic fermion-antifermion systems, the corresponding Dirac
sea is bottomless and can never be depleted, so the equivalence is exact even at the
nonperturbative level.
(An alternative operator approach to bosonization of a finite number of fermions in one 
space dimension is given in \cite{Dhar}.)

The phase space droplet approach works, in principle,
in any number of dimensions, at least semiclassically. 
In fact, an adaptation of this method properly taking into account the quantum
nature of the phase space was proposed as the starting
point for an {\it exact} bosonization in
arbitrary dimensions \cite{ncboz}. (For other approaches
to higher dimensional bosonization see \cite{hiboz}.) A full development
of this method and application to realistic, interacting higher dimensional
fermion systems remains an interesting project.

The success of the bosonization method relies on the fact that the two
systems share the same set of physical states and observables. This means that
many-body fermion operators and bosonic operators are into one-to-one
correspondence, and the two sets obey the same operator algebra.
More importantly, the Hilbert spaces on which these operators act in the two
systems are equivalent. This is crucial, as there are examples of systems with the
same algebra of observables but with their Hilbert spaces providing inequivalent
representations of this algebra. In fact, the albegra of hydrodynamic density operators
is such an example, as it admits several representations corresponding
to particles with inequivalent (and even exotic) statistics. It is important, therefore,
to have a realization of the algebra of observables that admits the many-body
fermionic Hilbert space as its (unique and irreducible) representation.

The droplet parametrization achieves just that.
Specifically, the phase space density $\hat \rho (x,p)$ is a `universal'
operator in terms of which all observables of the many-body system can be
expressed, including arbitrary (particle-number preserving) interactions.
The realization of this operator in terms of the operator representing the
boundary of the droplet leads to a representation of the full set of fermion
excitations and provides a complete `dictionary' between the two systems
\cite{alexios1}.

In this paper we push this method further in the tractable case of
one space dimension by deriving the bosonized form of the Hamiltonian for
a finite number of fermions for various cases (dispersion relations) that do not seem to
have been examined in the literature. This includes the standard results
of massless relativistic fermions and nonrelativistic fermions,
but generalizes them to arbitrary polynomials or analytic functions of
the momentum. Interactions are also considered, demonstrating
that the droplet method reveals interesting collective behavior around the Fermi sea.
This sets the stage for possible extensions to higher dimensions and
for potentially more realistic applications.

\section{Review of the droplet density method}

We consider $N$ fermions in one space dimension. For convenience, we
will assume space to be compact and will choose units such
that its periodicity be $2\pi$ (which quantizes single-particle momenta
in integer units).

The Weyl-ordered quantum many-body density operator is given by
\beq
\hat{\rho} (x,p) = {1 \over {(2\pi)^2}} \sum_{l = -\infty}^\infty \int dk \sum_{i=1}^{N} e^{i[l (\hat{x}_i-x)+k(\hat{p}_i-p])}
\label{qdo}
\eeq
Its Fourier transform $\hat{\rho} (l,k) = \sum_{i=1}^{N} e^{i(l \hat{x}_i+k\hat{p}_i)}$ satisfies the well-known ``sine" algebra \cite{sina}
\beq
[ \hat{\rho} (l,k)~,~ \hat{\rho} (l',k')] = -2 i \sin \bigl({{lk'-kl'} \over 2} \bigr) \hat{\rho} (l+l',k+k')
\label{sinealgebra}
\eeq
Given a single-particle Hamiltonian $H_{sp} (\hat{x}, \hat{p})$, expressed in a Weyl-ordered form in terms of $\hat{x},~\hat{p}$, one can write the exact quantum many-body Hamiltonian in terms of $\hat{\rho}(x,p)$ and $H_{sp} (x,p)$ as
\beq
\hat{H} = \sum _{i=1}^{N} H_{sp} (\hat{x}_i,\hat{p}_i) = \int H_{sp} (x,p) \hat{\rho}(x,p) dx dp
\label{manybodyH}
\eeq
This can be similarly extended to other quantum operators and also to cases with interactions.

In \cite{alexios1} it was shown that the quantum density operator $\hat{\rho}(x,p)$ can be explicitly written in terms of the boundary operator $\hat{R}(x)$, achieving the bosonization of any many-fermion system and further providing an explicit way of deriving the exact bosonized expressions for all quantum operators of the theory.

The quantum boundary field $\hat{R}(x)$ is a chiral field satisfying the commutation rule
\beq
[\hat{R}(x)~,~\hat{R}(x')] = -2 \pi i \delta ' (x-x')
\label{chiral}
\eeq
or in terms of its Fourier modes 
\beq
[\hat{R}_n~,~\hat{R}_{m}] = n \delta_{n+m}
\label{modes}
\eeq
where $\hat{R}(x) = \sum_{n=-\infty}^{+\infty} \hat{R}_n~e^{inx}$.
Semiclassically, it represents the value of the Fermi momentum at position $x$. In fact, there are
two such mutually commuting chiral fields, ${\hat R}(x)$ and $\hat{{\bar R}}(x)$, 
corresponding to the two Fermi momenta of a finite particle distribution. Particle excitations around each Fermi
momentum actually factorize, so for most purposes considering each section separately will suffice.
In the following we will consider only one chiral sector generated by ${\hat R}(x)$. For that purpose we use phase space coordinates $(x,p) \in [-\pi, \pi] \times [0, \infty)$, where the momentum $p$ is positive, essentially considering only ``half" the Fermi sea states. We
will return to the issue of considering both sectors in the section on interactions.

The Fock space on which the $R_n$'s act consists of the vacuum state $|0\rangle$, which is annihilated by the positive modes $\hat{R}_n~(n > 0)$ and the excited states generated by the action of the negative modes, $\hat{R}_{-n}$, on the vacuum. Since $\hat{R}_0$ commutes with all $\hat{R}_n$'s, the action of $\hat{R}_0$ on the vacuum defines a conserved quantity $N$, which can be identified with the number of fermions, $\hat{R}_0 |0 \rangle = {N \over 2} |0 \rangle$. (The remaining half
are assigned to the zero mode $\hat{\bar R}_0$ of the other chiral sector.)

In \cite{alexios1} it was shown that the Fourier transform of the normally ordered quantum density operator can be written in terms of the chiral field $\hat{R}(x)$ as
\beq
\hat{\rho}(l,k) = \int {1 \over{ 4\pi i \sin {k \over 2}}} e^{a(x,k)} e^{b(x,k)} e^{ik \hat{R}_0} e^{ilx} dx
\label{rhoR}
\eeq
where
\beqar
a(x,k) &=& \sum_{n>0} {2i \over n} \sin {{nk} \over 2} \hat{R}_{-n} e^{-inx} \nonumber \\
b(x,k) &=& \sum_{n>0} {2i \over n} \sin {{nk} \over 2} \hat{R}_{n} e^{inx}
\label{ab}
\eeqar
The above operators satisfy the commutation relations (\ref{sinealgebra}) when acting on the
Fock states of $R_n$.
It is straightforward to show using (\ref{rhoR}, \ref{ab}) that
\beq
\hat{\rho}(x,k) = {1 \over {2\pi}}  \sum_{l = -\infty}^\infty \hat{\rho}(l,k) e^{-ilx} = {:{e^{i \int_{x-{k \over 2}}^{x+{k \over 2}} \hat{R}(s) ds} :} \over {4\pi i \sin {k \over 2}}} 
\label{rhoxk}
\eeq
The small $k$-limit of the above expression is the Fourier transform of a step function, reproducing the semiclassical droplet result, $\hat{\rho}(x,p) = {1 \over {2\pi}}~ \theta (\hat{R}(x)-p)$.

The zero-mode contribution to the density operator is
\beq
\rho_0(x,p) = {1 \over {2\pi}} \int_{-\infty}^{\infty} {{e^{ik(R_0-p)}} \over {4 \pi i \sin {k \over 2}}} dk
\label{rho01}
\eeq
Evaluating this using a contour integration and a $k \rightarrow k-i\epsilon$ prescription, we find
\beq
\rho_0(x,p) = {1 \over {2\pi}} \theta (R_0-p) \sum_k \delta (R_0-p-{ 1 \over 2} -k)
\label{rho02}
\eeq
For a single-particle Hamiltonian of the form $H_{sp}(x,p) = h(p)$, the zero-point energy for the many-body system is
\beqar
H_0 &= &\int h(p) \rho_0(x,p) dx dp = \int_0^{R_0} h(p) \sum_k \delta (R_0-p-{ 1 \over 2} -k) ~dp \nonumber \\
&=& \sum_{n=\epsilon}^{R_0-{1 \over 2}} h(n)
\label{H0}
\eeqar
as expected, where $\epsilon = 0$ ($\half$) if $N= 2R_0$ is odd (even) respectively.

\section{Derivation of Hamiltonians for polynomial dispersion}

Let us first consider the many-body quantum Hamiltonian corresponding to the single-particle Hamiltonian of the form $H_{sp}(x,p) = p^n$.

The zero-mode contribution to the energy is (for simplicity we are going to consider $N$ odd)
\beq
H_0^{(n)} = \int p^n \rho_0 (x,p) dx dp = \sum_{l=0}^{R_0-{1 \over 2}} l^n
\label{H0n}
\eeq
The sum in (\ref{H0n}) has a compact expression in terms of Bernoulli polynomials as follows,
\beq
H_0^{(n)} = \sum_{l=0}^{R_0-{1 \over 2}} l^n = {{B_{n+1}(R_0+ {1 \over 2}) - B_{n+1}} \over {n+1}}
\label{H0B}
\eeq
where $B_{n+1}(0)=B_{n+1}$ are the corresponding Bernoulli numbers.
For large $R_0$, the zero-point energy in (\ref{H0n}, \ref{H0B}) reduces to the semiclassical droplet result $H_0^{(n)} \rightarrow {{R_0^{n+1}} / {(n+1)}}$.

Using (\ref{rhoxk}, \ref{rho01}) we find that the quantum many-body Hamiltonian characterizing the excitations of the system can be written as
\beqar
H^{(n)}-H_0^{(n)} &=& {1 \over {2\pi}} \int p^n ~{:{e^{i \int_{x-{k \over 2}}^{x+{k \over 2}} \hat{R}(s) ds} : - e^{ik R_0}} \over {4\pi i \sin {k \over 2}}} e^{-ikp} ~dk ~dp~dx \nonumber \\
&=& \int_{-\pi}^{\pi} dx \Bigl[ \Bigl({{-i \del_k} \over 2}\bigr)^n \bigl( {:{e^{i \int_{x-k}^{x+k} \hat{R}(s) ds} : - e^{2ik R_0}} \over {4\pi i \sin k }} \bigr) \Bigr] _{k=0}
\label{dH} 
\eeqar
The exponential term containing the boundary field $R(s)$ can be written as a series expansion in terms of $R(x)$ and its even derivatives as (we drop hats from operators from now on)
\beq
:\exp{\Bigl(i \int_{x-k}^{x+k} R(s) ds}\Bigr): = :\exp{~\Bigl(2 i k \Bigl[ R(x) +\sum_{l=1} k^{2l} {{\del_x^{2l}R(x)} \over {(2l+1)!}} \Bigr] \Bigr) \equiv :\exp \Bigl( 2 i k ( R(x) +R_D (k,x))}\Bigr):
 \label{expansion}
\eeq
where the term $R_D(k,x)$ contains only derivatives of the field $R(x)$. As a result, the Hamiltonian in (\ref{dH}) can be written as an integral of a polynomial given in terms of the boundary field $R(x)$ and its even derivatives. 

Using (\ref{expansion}), we can write the expression for the Hamiltonian in (\ref{dH}) as 
\beq
H^{(n)}-H_0^{(n)} = \Bigl(-{i \over 2}\Bigr)^n \int_{-\pi}^{\pi} dx ~ (\del_k)^n \bigl[ :{{e^{2i k R(x)} - e^{2i k R_0}}  \over {4\pi i\sin k }} +
{{e^{2i k (R(x) + R_D(x,k))} -e^{2i k R(x)} } \over {4\pi i\sin k }}  : \bigr] _{k=0}
\label{dH1}
\eeq
The first term in (\ref{dH1}) is a polynomial in $R(x)$ (does not contain any derivatives of the field), while the second term is an expression containing even derivatives of the field. We shall now calculate these two terms separately. A useful formula for these calculations is the one giving the generating function for Bernoulli polynomials, namely
\beq
{{t e^{zt}} \over {e^t -1}} = \sum_{n=0} ^ {\infty} B_n (z) {t^n \over {n!}}
\label{bernoulli}
\eeq
Using (\ref{bernoulli}) we find 
\beq
:{{e^{2i k R(x)} - e^{2i k R_0}}  \over {\sin k }}: = 2i \sum_{l=0}^{\infty} : {{[ B_{l+1}(R + {1\over 2}) - B_{l+1} (R_0 + {1\over 2}) ]} \over {l+1}} : {{(2ik)^l } \over {l!}}
\label{nd}
\eeq
where we used the fact that $B_0(z)=1$ for any $z$. Using (\ref{nd}) in (\ref{dH1}) we easily find that
\beq
H^{(n)}-H_0^{(n)} = {1 \over {2\pi}} \int dx~ \bigl[ : {{B_{n+1}(R(x) + {1\over 2}) - B_{n+1} (R_0 + {1\over 2}) } \over {n+1}}  + ({\rm derivative~terms}): \bigr]
\label{ndH}
\eeq
Further using (\ref{H0n}) we find that
\beq
H^{(n)} = {1 \over {2\pi}} \int dx~ \bigl[ : {{B_{n+1}(R(x) + {1\over 2}) - B_{n+1} } \over {n+1}}  + ({\rm derivative~terms}) :\bigr]
\label{ndH1}
\eeq
We now outline the calculation of the derivative terms in (\ref{ndH1}) which is somewhat more involved. Using the following relation for Bernoulli polynomials
\beq
B_s (x+y) = \sum_{j=0}^s ~{{s!} \over {j! (s-j)!}} ~y^j ~B_{s-j} (x)
\label{bernoulli1}
\eeq
and (\ref{bernoulli}) we find that
\beq
:{{e^{2i k (R(x) + R_D(x,k))} -e^{2i k R(x) }} \over {\sin k }} :=  \sum_{s=1}^{\infty} \sum_{j=1}^{s}~ {1 \over {j! (s-j)!}} ~: B_{s-j} \bigl( R+ {1 \over 2} \bigr) ~(R_D)^j : ~(2ik)^{s-1}
\label{dr}
\eeq
where $(R_D)^j$ can be expanded in terms of even derivatives of $R(x)$ as
\beqar
(R_D(k,x))^j & = & \bigl( \sum_{i=1}^{\infty} k^{2i}d_{2i} \bigr)^j = \sum_{r_1+ \cdots r_i=j} {{j!} \over {r_1!r_2!\cdots r_i!}} k ^{2r_1+4r_2+\cdots 2ir_i} d_2^{r_1} d_4^{r_2} \cdots d_{2i}^{r_i} \nonumber \\
d_{2i} & \equiv & {{\del_x^{2i} R(x)} \over {(2i+1)!}}
\label{dr1}
\eeqar
Defining the terms in the Hamiltonian (\ref{ndH1}) that depend on the derivatives of the field $R(x)$ as $H_D^{(n)}$ and using (\ref{dr},\ref{dr1}) we find
\beq
H_D^{(n)} = {1 \over {2\pi}} \int dx ~:~\sum_{l=1}^{\{n/2\}}\sum_{j=1}^{l} {{(-1)^l} \over {2^{2l}}} {{n! ~B_{n+1-2l-j}\bigl(R + \half)} \over {(n+1-2l-j)!}} \sum_{\substack{r_1+ \cdots r_i=j, \\ r_1+2r_2 + \cdots + ir_i=l}} {{d_2^{r_1} d_4^{r_2} \cdots d_{2i}^{r_i}} \over {r_1!r_2!\cdots r_i!}} ~:
\label{DHn}
\eeq
where 
$\{n/2\}= n/2$ if $n$ is even ($= {{n-1} \over 2}$ if $n$ is odd). This gives an expansion in terms of the total number of derivatives.

Expressions (\ref{ndH1}, \ref{DHn}) give the exact quantum many-body Hamiltonian for a fermionic system with single particle Hamiltonian of the form $p^n$ in terms of the boundary field $R(x)$ and its derivatives. Here are the results for $H^{(n)}~,n=1,\cdots 5$:
\vskip .2in
\noindent
\underline{$n=1$}
\beq
H^{(1)} = {1 \over {2\pi}} \int dx~: {{B_2 \bigl(R + \half \bigr) -B_2} \over 2}: ~ =  {1 \over {2\pi}} \int dx~: \bigl( {{R^2} \over 2} -{1 \over 8} \bigr):
\label{H1}
\eeq

\vskip .2in
\noindent
\underline{$n=2$}
\beq
H^{(2)}  = {1 \over {2\pi}} \int dx~: {{B_3 \bigl(R + \half \bigr) -B_3} \over 3}  :  ~= {1 \over {2\pi}} \int dx~: \bigl( {{R^3} \over 3} -{{R} \over 12} \bigr):
\label{H2}
\eeq

\vskip .2in
\noindent
\underline{$n=3$}
\beqar
H^{(3)} & = &{1 \over {2\pi}} \int dx~: {{B_4 \bigl(R + \half \bigr) -B_4} \over 4} - {{B_1 \bigl( R + \half\bigr)} \over 4} {R''(x)} :  \nonumber \\
& = & {1 \over {2\pi}} \int dx~: \bigl( {{(R^2 -{ 1 \over 4})^2} \over 4} -{{R R'' } \over 4}  \bigr):
\label{H3}
\eeqar

\vskip .2in
\noindent
\underline{$n=4$}
\beqar
H^{(4)} & = &{1 \over {2\pi}} \int dx~: {{B_5 \bigl(R + \half \bigr) -B_5} \over 5} - {1 \over 2} {B_2 \bigl( R + \half\bigr)} {R''(x)}  :\nonumber \\
& = & {1 \over {2\pi}} \int dx~: \bigl( {{R^5} \over 5} -{{R^3} \over 6} + {{7R} \over 240}-{{R^2 R'' } \over 2}  \bigr):
\label{H4}
\eeqar

\vskip .2in
\noindent
\underline{$n=5$}
\beqar
H^{(5)} & = &{1 \over {2\pi}} \int dx~: {{B_6 \bigl(R + \half \bigr) -B_6} \over 5} - {5 \over 6} {B_3 \bigl( R + \half\bigr)} {R''(x)} +{ 1 \over 16} {B_1 \bigl( R + \half\bigr)} {R''''(x)}  +{ 5 \over 48} (R'')^2 :\nonumber \\
& = & {1 \over {2\pi}} \int dx~: \bigl( {{R^6} \over 6} -{{5 R^4} \over 24}+{{7 R^2} \over 96} -{1 \over 128}-{5  \over 6} (R^3-{{R} \over 4})R'' +{{ R R'''' } \over 16} +{{5 ( R'' )^2} \over 48}\bigr):
\label{H5}
\eeqar
In all the above expressions we have neglected total derivative terms.
\section{Extension to arbitrary functions of the momentum}

The calculation of the many-body quantum Hamiltonian can now be extended to cases where the single-particle Hamiltonian is an arbitrary function of momentum, namely $H_{sp}=f(p)$.

Suppose that $f(p)$ admits a Taylor series expansion
\beq
f(p) = \sum_{n=0}^{\infty} {{p^n} \over {n!}}~ \del^nf(0) 
\label{taylor}
\eeq
Then
\beq
H= \sum_{n=0}^{\infty} {{\del^n f (0)} \over {n!}} \int p^n \hat{\rho} (x,p) dx dp = \sum_{n=0}^{\infty} {{\del^n f (0)} \over {n!}} H^{(n)}
\eeq
For simplicity we shall split the calculation of $H$ into two terms: $H_{ND}$ will be the contribution from the non-derivative part of $H^{(n)}$, given in (\ref{ndH1}), and $H_D$ will be the contribution from the derivative part $H_D^{(n)}$, given in (\ref{DHn}):
\beqar
H &=& H_{ND} + H_D \nonumber \\
H_{ND} & = & { 1 \over {2\pi}} \int dx~ \sum_{n=0}^{\infty} {{\del^n f (0)} \over {n!}}  : {{B_{n+1}(R(x) + {1\over 2}) - B_{n+1} } \over {n+1}} :   \label{H} \\
H_D & =& { 1 \over {2\pi}} \int dx~ \sum_{n=0}^{\infty} {{\del^n f (0)} \over {n!}} :~\sum_{l=1}^{\{n/2\}}\sum_{j=1}^{l} {{(-1)^l} \over {2^{2l}}} {{n! ~B_{n+1-2l-j}\bigl(R + \half)} \over {(n+1-2l-j)!}} \sum_{\substack{r_1+ \cdots r_i=j, \\ r_1 + \cdots + ir_i=l}} {{d_2^{r_1} d_4^{r_2} \cdots d_{2i}^{r_i}} \over {r_1!r_2!\cdots r_i!}} ~: \nonumber
\eeqar
Using the following relation for Bernoulli polynomials 
\beq
\int_a^x B_r(t) = {{B_{r+1}(x) - B_{r+1}(a) } \over {r+1}} 
\label{bernoulli2}
\eeq
we can write
\beq
H_{ND}= { 1 \over {2\pi}} \int dx~ : \int_0^{R+ {1 \over 2}} \sum_{n=0}^{\infty} {{B_n(t)} \over {n!}}  \del^n f (0): 
\label{HH1}
\eeq
Further the Bernoulli polynomial can be expanded in terms of Bernoulli numbers as
\beq
B_n(t) = \sum_{k=0}^{n} {{B_k} \over {k!}} {{n!} \over {(n-k)!}} t^{n-k} = \sum_{k=0}^{n} {{B_k} \over {k!}} (\del_t)^k t^n
\label{bernoulli3}
\eeq
Using $z=0$ and replacing $t \rightarrow \del_t$ in (\ref{bernoulli}) one can easily show that
\beq
B_n(t) = {{\del} \over {e^{\del} -1}} t^n
\label{bernoulli4}
\eeq
Using this in (\ref{H1}) we find that we can express $H_{ND}$ as
\beq
H_{ND} = { 1 \over {2\pi}} \int dx~ : \int_0^{R+ {1 \over 2}} dt~{{\del} \over {e^{\del} -1}} f(t):
\label{1H1}
\eeq
If we define $G(t)= {{\del} \over {e^{\del} -1}} f(t)$ then $G(t)$ has to satisfy the relation $G(t+1)-G(t)=\del f(t)$. For special cases of $f(t)$, $G(t)$ has a compact form. For example,
\beqar
f(t)= t^n & \Longrightarrow & G(t)= B_n(t) \nonumber \\
f(t)= e^{at} & \Longrightarrow & G(t)={{e^{at}} \over {e^a-1}} \nonumber
\eeqar
In general though, 
\beq
G(t) = {{\del} \over {e^{\del} -1}} f(t) = f(t) + B_1 \del f(t) + \sum_{k=1}^{\infty} {{B_{2k}} \over {(2k)!}} \del ^{2k} f(t)
\label{Gt}
\eeq
Using this in equation (\ref{1H1}) we find
\beqar
H_{ND}  \!\!\!\!\! &=&\!\!\!\!\! \int {dx \over {2 \pi}}: \Bigl[ \int_0^{R + \half} dt f(t) - {1 \over 2} \Bigl( f(R + \half) - f(0) \Bigr) + \sum_{k=1}^{\infty} {{B_{2k}} \over {(2k)!}}[ \del_u ^{2k-1}f(u)|_{u=R + \half} - \del _u^{2k-1} f(u)|_{u=0}] \Bigr] : \nonumber\\
&\equiv &\!\!\!\!\!  \int {dx \over {2 \pi}}: F\bigl( R(x) + \half \bigr):
\label{2H1}
\eeqar
where we used the fact that $B_1= -1/2$. The expression (\ref{2H1}) is not surprising. When $H_{ND}$ acts on the vacuum state, $R + \half$ is replaced by $(N+1)/2$ which is an integer (we have assumed $N$ to be odd). In this case (\ref{2H1}) is just the Euler-MacLauren formula connecting a sum with an integral, namely
\beqar
E & = & \sum_{i=0}^{{{N-1}\over 2}} f(i)  \\
&=& \int_0^{{{N+1} \over 2}} dt f(t) - {1 \over 2} \Bigl( f({{N+1} \over 2}) - f(0) \Bigr) + \sum_{k=1}^{\infty} {{B_{2k}} \over {(2k)!}}[ \del_u ^{2k-1}f(u)|_{u={{N+1} \over 2}} - \del _u^{2k-1} f(u)|_{u=0}] \nonumber
\label{E}
\eeqar
as expected. One can view the operator expression $: F\bigl( R(x) + {1 \over 2} \bigr):$ in (\ref{2H1}) as the ``operator" definition of (\ref{E}) when $R_0=N/2$ is replaced by the field $R(x)$. 

Let us now evaluate the part of the Hamiltonian that contains derivatives of the boundary field. We found in (\ref{DHn}) that the $2l$-derivative term in the $H^{(n)}$ case ($n \ge 2l$) is of the form
\beq
H_{D^{2l}}^{(n)} = {1 \over {2\pi}} \int dx ~:~\sum_{j=1}^{l} {{(-1)^l} \over {2^{2l}}} {{n! ~B_{n+1-2l-j}\bigl(R + \half)} \over {(n+1-2l-j)!}} \sum_{\substack{r_1+ \cdots r_i=j, \\ r_1+2r_2 + \cdots + ir_i=l}} {{d_2^{r_1} d_4^{r_2} \cdots d_{2i}^{r_i}} \over {r_1!r_2!\cdots r_i!}} ~:
\label{DHn2l}
\eeq
Using the following property of the Bernoulli polynomials
\beq
\del_t^i B_n(t) = {{n!} \over {(n-i)!}} B_{n-i}(t) ~~~,~~~n\ge i
\label{bernoulli4}
\eeq
as well as equations (\ref{bernoulli3}-\ref{DHn2l}) one can show that the $2l$-derivative part of $H_D$ in (\ref{H}) can be written as 
\beq
H_{D^{2l}} = {1 \over {2\pi}} \int dx ~:~\sum_{j=1}^{l} {{(-1)^l} \over {2^{2l}}}  \sum_{\substack{r_1+ \cdots r_i=j, \\ r_1+2r_2 + \cdots + ir_i=l}} {{d_2^{r_1} d_4^{r_2} \cdots d_{2i}^{r_i}} \over {r_1!r_2!\cdots r_i!}} \del_u^{2l+j} F(u)|_{u=R + \half} ~:
\label{DH}
\eeq
Combining the non-derivative term, eq. (\ref{2H1}), and the derivative terms, eq.(\ref{DH}), we find that the full Hamiltonian can be written as
\beq
H = {1 \over {2\pi}} \int dx ~:\exp \Bigl( ~\sum_{i=1}^{\infty} {{(-1)^i \del_x^{2i}}R(x) \over {2^{2i} (2i+1)!}}  \del_u^{2i+1} \Bigr)~ F(u)|_{u=R + \half}~:
\label{Hf}
\eeq
where $F\bigl( R + \half \bigr)$ is defined in (\ref{2H1}).
\vskip .2in

In the case where the single particle Hamiltonian is polynomial in momentum, $f(p)=p^n$, we have 
\beq
F(R+{1 \over 2})  = \int _0^{R+ {1 \over 2}} dt~ B_n(t) = {{B_{n+1}(R(x) + {1\over 2}) - B_{n+1} } \over {n+1}} 
\eeq
and the corresponding expression (\ref{Hf}) provides a compact expression for (\ref{ndH1}, \ref{DHn}).

An interesting application of (\ref{Hf}) is in the case of noninteracting massive relativistic fermions, where $f(p) = \sqrt{p^2+m^2}$.
(We consider them spinless, or assume spin to be `frozen' in one direction.)
In this case $F(R+{1 \over 2})$, as defined in (\ref{2H1}), has an infinite series expansion in $1/N$. Keeping up to $\O(1)$ terms, we find
\beq
H = \int {{dx} \over {2\pi}} :F(R(x) +\half):
\eeq
where
\beqar
F(R+\half) = &&\half (R+\half) \sqrt{m^2 + (R+\half)^2} + {m^2 \over 2} \ln \left( {{R+\half} \over m} + \sqrt{1+ {{(R+\half)^2} \over {m^2}}} ~\right) \nonumber \\
&& -\half \left( \sqrt{m^2 + (R+\half)^2} -m \right) +{1 \over {12}} \left( {{R+\half} \over {\sqrt{m^2 + (R+\half)^2}}}
- {\epsilon \over {\sqrt{m^2 + \epsilon^2}}} \right)
\label{fr}
\eeqar
where $\epsilon \rightarrow 0$ eventually. (Keeping $\epsilon$ finite is important in considering the $m \rightarrow 0$ limit.) Terms containing derivatives of the boundary field are $\O (1/N)$ and are neglected in (\ref{fr}).  

The $m \rightarrow 0$ of the Hamiltonian in (\ref{fr}) reproduces the expression for massless relativistic particle in (\ref{H1}).

\section {Interactions}

The droplet formalism allows for the introduction of interactions in a straightforward way.
We shall consider space-dependent interactions only, and focus on one-body and two-body
potentials.

The important element when including interactions is that, in general, we may not
consider each chiral sector of the theory separately. Indeed, the space particle density operator
\beq
{ \rho} (x) = \int dp~ { \rho} (x,p) ~, ~~~ [{ \rho} (x) ~,~ { \rho} (x') ] = 0
\eeq
would not even commute with itself if only one sector were included. We thus consider
the two boundary operators ${ R}(x)$ and ${ {\bar R}}(x)$ corresponding to the two
Fermi momenta in semiclassical phase space. The opposite-chirality boundary field $\bar R$
satisfies 
\beq
[\bar{R}(x)~,~\bar{R}(x')] = +2 \pi i \delta ' (x-x')
\label{antichiral}
\eeq
with a crucial opposite sign in the commutator. The modes are, then, defined with a
corresponding negative sign
\beq
{\bar R} (x) = \sum_{n=-\infty}^{\infty} {\bar R}_n ~ e^{-inx}
\eeq
so that positive (negative) $n$ still label annihilation (creation) operators. The modes satisfy
\beq
[ {\bar R}_n ~,~ {\bar R}_m ] = n ~\delta_{n+m} ~,~~~~ 
[{\bar R}_n  ~,~ R_m ] = 0 
\eeq
All corresponding many-boby operators, such as the density operator and
the Hamiltonian, become the {\it difference} of two expressions involving
$R$ and $\bar R$. The space density, in particular, becomes
\beq
\rho (x) = \frac{1}{2\pi} [ R(x) - {\bar R} (x) ]
\eeq
and in terms of modes
\beq
\rho_n = \frac{1}{2\pi} ( R_n - {\bar R}_{-n} )
\label{rhoRR}
\eeq
We see that the $\rho_n$ commute among themselves. The total
number of particles is
\beq
N = 2\pi \rho_0 = R_0 - {\bar R}_0
\eeq
and in a symmetric situation $R_0 = -{\bar R}_0 = \frac{N}{2}$.

\subsection{One-body potentials}
In the presence of a one-body potentail $V(x)$ the Hamiltonian aquires the
extra term
\beq
U = \int dx V(x) \rho(x) = \sum_{n=-\infty}^{\infty} V_{-n} R_n - 
\sum_{n=-\infty}^{\infty} V_n {\bar R}_n
\eeq
and the full Hamiltonian, consisting of a general momentum term as analyzed in the
previous sections and the potential term, becomes
\beq
H_{int} = H[R] + \int dx V(x) R(x) - H[{\bar R}] - \int dx V(-x) {\bar R}(x)
\eeq
with $H[R]$ the free Hamiltonian as given in ({\ref{Hf}). 
We see that in the case of one-body potentials the full Hamiltonian
remains the sum of two chiral terms, so the two sectors do not mix and can be
considered separately.

To explore the spectrum of the above Hamiltonian, we will work to leading
order in $1/N$. All the derivative terms in (\ref{Hf}) are subleading, so we
get for the $R$-chirality sector
\beq
H = \frac{1}{2\pi} \int dx : F_o \left( R(x) \right) + V(x) R(x) :
\label{57}
\eeq
where we defined $F_o (R) = F(R+\half )$. Note that for a `relativistic'
single-particle Hamiltonian linear in momentum ($F_o$ quadratic) the
leading $1/N$ approximation is (perturbatively) exact.

We proceed to linearize and
diagonalize the Hamiltonian by considering the solution $R_c (x)$ 
to the classical equation
\beq
F'_o (R_c (x) ) + V(x) = E_F
\label{Efermi}
\eeq
$E_F$, which can be thought of as the ``chemical potential" Lagrange multiplier
enforcing the constraint $R_0 - {\bar R}_0 = N$, is essentially the semiclassical Fermi
level energy. It is determined by the requirement that the solution of the above
equation satisfy
\beq
\frac{1}{2\pi} \int dx \, R_c (x) = \frac{N}{2}
\eeq
assuming similar equations for $\bar R$ and a symmetric situation.
Writing
\beq
R(x) = R_c (x) + r(x)
\eeq
the new operator $r(x)$ satisfies the same chiral boundary algebra as $R(x)$
as it is shifted by a c-number term. It terms of $r(x)$ the Hamiltonian becomes
(dropping explicit $x$-dependence)
\beq
H = \frac{1}{2\pi} \int dx : \left\{ F_o (R_c ) + V R_c +
\left[ F'_o (R_c ) + V \right] r + \half F''_o (R_c ) r^2
+ O (r^3 ) \right\} :
\label{Hr3}
\eeq
Due to the equation (\ref{Efermi}) the term in the middle bracket is
a constant and it contributes to the Hamiltonian a term proportional
to $E_F ~r_0$. The zero mode of $r$ corresponds to additional particles
at this Fermi level. Since $R_0 - {\bar R}_0 = N$, any additional
particle appearing in $R_0$ would have to come by migrating from the
opposite Fermi level ${\bar R}_0$. Such moves across Fermi levels
constitute nonperturbative excitations and, at any rate, can be accounted
for separately as $r_0 = {\bar r}_0$ are Casimirs and commute with
all other operators. Omitting this term, then, the Hamiltonian to leading
order in $1/N$ (quadratic in $r$) becomes
\beq
H = E_o + \frac{1}{2\pi} \int dx ~\frac{1}{2} : v r^2 :
\eeq
with $v (x)  = F_o'' (R_c (x) )$ the semiclassical Fermi velocity and 
$E_o$ the constant term in (\ref{Hr3}). We define a new coordinate $y$ by
\beq
\frac{dx}{v(x)} = \alpha \, dy~,~~~ 2\pi \alpha = \int \frac{dx}{v(x)}
\eeq
with the coefficient $\alpha$ chosen so that $y$ have a periodicity of $2\pi$.
We also define the new operator $\phi = v r$ whose Fourier modes in the
$y$ variable are
\beq
\phi_n = \frac{1}{2\pi} \int dy \, \phi(y) \, e^{i n y} 
=  \frac{1}{2\pi} \int dx~ r(x) \, e^{i n y(x)}
\eeq
and satisfy the standard chiral algebra
\beq
[ \phi_n \, ,\, \phi_m ] = n~ \delta_{n+m}
\eeq
In terms of the above, the Hamiltonian becomes (omitting the constant $E_o$)
\beq
H = \frac{1}{2\pi} \int dy \frac{1}{2} \alpha : \phi(y)^2 : ~
= \alpha \sum_{n>0}^{\infty} ~\phi_{-n} \phi_n 
\eeq
This is a set of decoupled harmonic oscillators. The normal ordering in the
new field obviously has to be done in terms of the modes $\phi_n$ leading
to a positive definite Hamiltonian. The excitations are equidistant over the
ground state with a spacing $\alpha$ between the single-particle energy levels.

The above results can be recovered in the many-body language through a
semiclassical calculation. The new coordinate $y$ is essentially the
time of flight of a particle at the Fermi energy $E_F$ and $\alpha$ is the
single-particle energy spectrum spacing near the Fermi level as derived
in a WKB approximation. What the above calculation shows is that the
WKB result becomes exact in the large-$N$ limit, and it is
exact to {\it all} orders in $1/N$ for a single-body kinetic term
linear in the momentum.

Using the above formalism we can also keep higher order terms in $r$
and $1/N$ and calculate corrections to the many-body energy spectrum. In this case, $F_o(R)$ in (\ref{57}) has to be modified by keeping the appropriate subleading derivative terms appearing in (\ref{Hf}). 
For arbitrary potentials, however, this becomes a rather complicated
calculation.

\subsection{Two-body potentials}

The situation becomes a lot more interesting when we consider two-body
potentials. The two chiral sectors mix, and genuine collective effects
come into play.

We will consider translationally invariant, symmetric two-body potentials
of the form
\beq
U = \sum_{i<j} V( x_i - x_j )
\eeq
with $V(x)$ an even real function of $x$. We will also assume that $V(x)$
is regular at $x=0$, so no singularities arise as two particles coincide.
The case of singular potentials, such as the Coulomb ($V \sim |x|^{-1}$)
or the ``Calogero" ($V \sim x^{-2}$) potential, require a more careful
treatment.

The potential energy can be expressed in terms of the particle density
operator $\rho (x)$ as
\beq
U = \half \left[ \int dx dy \, V(x-y) \, \rho (x) \, \rho (y) - \int dx V(0) \rho (x)
\right]
\eeq
where the second term removes the particle self-interactions (the terms
$i=j$ in the many-body sum). Expressing $V(x)$ in terms of its Fourier
modes
\beq
V(x) = \sum_n V_n e^{inx}
\eeq
with $V_{-n} = V_n$ real, the above expression becomes
\beq
U = (2\pi )^2 \sum_{n>0} V_n \rho_{-n} \rho_n +
\half (2\pi )^2 V_0 \rho_0^2 - \pi V(0) \rho_0
\eeq
Expressing the density in terms of chiral modes as in (\ref{rhoRR})
and assuming again a symmetric situation in which $R_0 = -{\bar R}_0
= \frac{N}{2}$ we obtain
\beq
U = \half V_0 N (N-1) + \sum_{n>0} (n-N) V_n +
\sum_{n>0} V_n \left(R_{-n} R_n + {\bar R}_{-n} {\bar R}_n
- R_n {\bar R}_n - R_{-n} {\bar R}_{-n} \right)
\label{Ud}
\eeq
where we also used $V(0) = V_0 + 2 \sum_{n>0} V_n$.

Before proceeding to the full Hamiltonian, we comment on the validity range
of the above expression for the interaction energy. As stated earlier,
droplet bosonization results are perturbatively exact, up to the point that
large excitations deplete the Fermi sea and mix the two chiral sectors.
The Fourier mode $V_n$ of the interaction potential, on the other hand,
generically creates
excitations of order $n$ in the fermion state. To prevent the above-mentioned
nonperturbative effects, $n$ should be of order smaller than $N$. Therefore,
only smooth enough potentials, with vanishing Fourier modes as $n$ becomes
of order $N$, are reliably expressed in the droplet bosonization formula (\ref{Ud}).

It is instructive to consider the special case of a delta-function two-body
potential
\beq
V(x) = c \, \delta (x) ~,~~~ V_n = \frac{c}{2\pi}
\eeq
This potential is special in that it clearly violates the condition stated above,
as it has nonvanishing modes for all $n$, and in that it is actually irrelevant
for fermions. Indeed, its support is only at particle coincidence points where
the fermionic wavefunction vanishes. Therefore, its expectation value for
any fermionic state should vanish.

Applying formula (\ref{Ud}) for $V_n = c/2\pi$ and for the free fermion
homogeneous ground state annihilated by all $R_n$ and ${\bar R}_n$, $n>0$,
the expectation value of the operator term vanishes. The constant
terms give a formally infinite result due to the infinite
sum in $V_n$. The distance between the two Fermi levels in this state, however,
equals $N-1$, so only modes with $n$ up to $N-1$ should be considered. Truncating
the sum over such modes we have
\beq
\left< U \right> =  \half \frac{c}{2\pi} N (N-1) + \sum_{n=1}^{N-1} (n-N)
\frac{c}{2\pi} = 0
\eeq
as required. For excited states, the operator part would also contribute.
For such states, modes with $n$ less than $N-1$ will also deplete the Fermi
sea and mix the two chiral sectors. The issue of projecting the Fock states
of the droplet oscillators to the proper Hilbert state of the fermion system
is quite nontrivial and will not be treated here.

We now proceed to deriving the full Hamiltonian and its spectrum. In the
absence of any potentials, the (free) ground state is the homogeneous
ground state of all oscillators $R_n$ and ${\bar R}_n$. We will
focus again to a symmetric situation and expand around the constant
background
\beq
R_c = -{\bar R}_c = \frac{N}{2}
\eeq
which amounts to setting the zero mode of the chiral fields to $N/2$.
Working, again, to leading order in $1/N$, the free Hamiltonian will be
\beq
H_o = K_o + \frac{1}{2\pi} \int dx \half v : R^2 + {\bar R}^2 :
\eeq
with $K_o$ the ground state kinetic energy and $v$ the Fermi velocity,
which is constant and common to both sectors:
\beq
K_o = 2 F_o \left(\frac{N}{2}\right) ~,~~~
v = F_o'' \left(\frac{N}{2}\right)
\eeq
The total Hamiltonian $H_o + U$ expressed in modes becomes
\beq
H = E_o + \sum_{n>0} (v+V_n ) (R_{-n} R_n + {\bar R}_{-n} {\bar R}_n )
- \sum_{n>0} V_n (R_n {\bar R}_n + R_{-n} {\bar R}_{-n} )
\eeq
with the constant term being
\beq
E_o = 2 F_o \left(\frac{N}{2}\right) + \half V_0 N(N-1) +\sum_{n>0}
(n-N) V_n
\eeq
We see that the two chiral sectors mix, but individual modes decouple.
We focus, therefore, on the operator part of the Hamiltonian for a single
mode $n$. Each such term can be brought to standard oscillator form
through a Bogoliubov transformation. To do that, we first decouple the two
sectors by performing the redefinitions
\beq
A_n = \frac{1}{\sqrt 2} ( R_n + {\bar R}_n ) ~,~~~
B_n = \frac{1}{\sqrt 2} ( R_n - {\bar R}_n )
\eeq
The Hamiltonian for mode $n$ becomes
\beq
H_n = (v+V_n) A_{-n} A_n - \half V_n (A_n^2 + A_{-n}^2)
+ (v+V_n) B_{-n} B_n +\half V_n (B_n^2 + B_{-n}^2)
\eeq
The $A$ and $B$ parts are now decoupled and essentially identical in
form up to a sign (which can be absorbed by the redefinition
$B_n \to i\, {\rm sgn}(n) B_n$). The $A$ part can be diagonalized in the standard way
by defining the new oscillator operators
\beq
a_n = \ch \theta \, A_n - \sh \theta \, A_{-n}
\eeq
The new $a_n$ obey the same algebra as $R_n$. With the
choice of the parameter $\theta$
\beq
\th 2\theta = \frac{V_n}{v+V_n} ~~~~{\rm or} ~~~~
\ch \theta = \sqrt{\frac{v+V_n}{2 \sqrt{v(v+2V_n )}} + \half}
\label{theta}
\eeq
the $A$ part of the Hamiltonian takes the form
\beq
H(a_n ) = \sqrt{v(v+2V_n )} \, a_{-n} a_n - \frac{n}{2} \left( v+V_n 
- \sqrt{v(v+2V_n )} \right)
\eeq
The $B$ part of the Hamiltonian can be diagonalized with a similar
transformation involving the opposite parameter $\theta$:
\beq
b_n = \ch \theta \, B_n + \sh \theta \, B_{-n}
\eeq
Putting everything together we obtain the final, decoupled, normal
form for the full Hamiltonian
\beq
H = E_{G} + \sum_{n>0} \sqrt{v(v+2V_n )} \,  ( a_{-n} a_n + b_{-n} b_n )
\label{Hbogo}
\eeq
with the ground state energy
\beq
E_G = 2 F_o \left(\frac{N}{2}\right) + \half V_0 N(N-1) 
+\sum_{n>0} \left[ n \left(  \sqrt{v^2+2v V_n} - v \right)
- N V_n \right]
\eeq
The ground state is annihilated by the operators $a_n$ and $b_n$,
$n>0$, but it is actually a ``squeezed state" in terms of the old
oscillators $R_n$ and ${\bar R}_n$. Excited states are build as
Fock states of the oscillators $a_n$ and $b_n$, but with a renormalized
energy gap $\sqrt{v(v+2V_n )}$ instead of $v$. The gaps depend
on $V_n$ and are generically different for different modes, so the
degeneracy of the noninteracting particle spectrum is lifted. Interestingly,
the Hamiltonian is still the sum of two commuting pseudo-chiral sectors,
although each sector is really a mixture of the two chiral modes.

\subsection{Phase transition}

From the relation (\ref{theta}) it is clear that the transformation
leading to the above Hamiltonian (\ref{Hbogo}) is valid only if
\beq
\left| \frac{V_n}{v+V_n} \right| < 1 ~~~~ {\rm or} ~~~~
V_n > -\frac{v}{2}
\eeq
For $V_n$ below the above critical value the Hamiltonian becomes
unbounded from below. The homogeneous state over which this
Hamiltonian was constructed becomes unstable, signaling a
phase transition. In this case, higher order terms in $1/N$
(cubic and higher order in $R$ and $\bar R$) and nonperturbative
effects cannot be neglected any more.
The true ground state will involve lumping of particles together.

To gain some intuitive understanding of the above phase transition
we consider the case of regular nonrelativistic fermions of unit
mass and a
two-body potential with a single nonzero Fourier mode $V_1$.
The Fermi velocity is the same as the Fermi momentum
\beq
v = R_0 = \frac{N}{2}
\eeq
and the potential is
\beq
V(x) = 2 V_1 \cos x \simeq 2V_1 - V_1 x^2 + \cdots
\eeq
This potential is repulsive for $V_1 >0$, so clearly it cannot
induce any particle lumping and phase transitions, but it is
attractive for $V_1 <0$. For small particle separation it becomes
a mutual harmonic oscillator attraction. Setting $V_1 = -k/2$
for positive $k$, we can approximate the many-body Hamiltonian
for small separations as
\beq
H = \sum_i \half p_i^2 + \sum_{i<j} \half k (x_i - x_j )^2
= \sum_i \half p_i^2 + \sum_i \half Nk x_i^2 
- \half k \left( \sum_i x_i \right)^2
\eeq
So we have essentially $N$ fermions in an external harmonic
potential $\half Nk x^2$ but with the center of mass oscillator
energy removed. The zero-momentum ground state is
\beq
\Psi_0 = C \prod_{i<j} (x_i - x_j )~
e^{-\half \sqrt{Nk} \left[ \sum_i x_i^2 - \frac{1}{N}
\left(\sum_i x_i \right)^2 \right]}
\eeq
with $C$ a normalization constant. To leading order in $1/N$
this is the wavefunction of $N$ noninteracting fermions in a
harmonic trap. The density distribution in the ground state is the
well-known Wigner semicircle
\beq
\rho_0 (x) = \frac{\sqrt{Nk}}{\pi} \sqrt{a^2- x^2}
~,~~~ a = \left(\frac{4N}{k}\right)^\frac{1}{4}
\eeq
As long as the radius $a$ of the semicircle is much smaller than the
length $2\pi$ of the periodic space in which fermions live, the ground state
will be well approximated by the above wavefunction and fermions
will stay close to each other. That is, for particle separations larger
than $2a$ the fermion two-point function will vanish. This is the
``lumped" phase.

However, when the size of the semicircle becomes
comparable to or larger than the size of space, its ends will start
`touching'. Further, the quadratic approximation of the potential will
not be adequate any more. The fermions will assume a ground state
that fills the available space with a nonzero two-point function everywhere.
This is the ``uniform" phase. The transition will happen for
\beq
a \sim 1 ~~~{\rm or}~~~ k \sim N \sim v
\eeq
So for $V_1 = -k/2$ of order $-v$ a phase transition would occur.
Our droplet oscillator analysis sharpens this intuition into an exact
statement.

A modified argument works for a general mode $V_n$. The potential
now has $n$ minima on the circle. Close to each minimum the potential is
\beq
V\left(x+\frac{2\pi}{n}\right) = 2V_n \cos (nx) \simeq 2V_n - n^2 x^2 + \cdots
\eeq
For large negative $V_n = -k/2$ the energetically favorable configuration
for the ground state is for
the particles to distribute equally around $n$ equidistant points on
the circle, with $N/n$ particles around each point. Each particle still feels
a harmonic oscillator potential of strength $N n^2 k$, since all other particles
contribute to the potential irrespective on which of the $n$ points on the
circle they are. The distribution of particles around each of the $n$ points
is again a Wigner semicircle, but now with an oscillator strength $N n^2 k$ and a number
of particles $N/n$. The Wigner radius is, then,
\beq
a_n = \left( \frac{4N}{n^4 k} \right)^\frac{1}{4} = 
\frac{1}{n} \left( \frac{4N}{ k} \right)^\frac{1}{4}
\eeq
A phase transition will occur when the semicircles start `merging',
which will happen when their size becomes of order $2\pi/n$ (as there are
$n$ of them). That is, when
\beq
a_n \sim \frac{1}{n} ~~~ {\rm or} ~~~ k \sim N
\eeq
So we obtain the same critical value for $V_n$ as for $V_1$. In the presence of more
than one nonzero $V_n$, of course, the argument becomes more complicated.
The droplet analysis, however, shows that a phase transition occurs when any 
of the $V_n$ reaches the critical value $V_n = -v/2$.

It should also be clear that the above phase transition 
is an essentially nonperturbative
effect. Indeed, the zero-point fluctuations of $a_n$ and $b_n$
in the ground state are always of order $\sqrt n$, but the fluctuations
of $A_n$ and $B_n$ (and thus also of $R_n$ and ${\bar R}_n$) are
of order $\sqrt{n} \,\ch \theta$. As $V_n$ approaches $-v/2$, these fluctuations
diverge and become of order $N$. At that point, droplet
fluctuations become large enough to mix the two chiral components
and signal the onset of nonperturbative effects.

\section{Conclusions}

The droplet operator method is an intuitively appealing approach
to bosonization and, as demonstrated in the above analysis, can
be carried out to produce perturbatively exact results for finite $N$ and probe
interesting many-body physics.

An important open question is the applicability of similar methods
to higher dimensions. It was suggested in \cite{ncboz} that an
adaptation of this method into a `noncommutative' higher
dimensional chiral field would provide a perturbatively exact
bosonization in higher dimensions. The irreducible
representation space of this algebra was demonstrated to
reproduce the Hilbert space of the many-body fermion system,
and the states and energies of fermions in a two-dimensional
hermonic trap were correctly reproduced by the model.
What is still missing is an exact mapping of operators 
that would provide a complete dictionary
between the two descriptions. That is, a droplet expression for the
density operator, which is a universal operator of the system,
is lacking. A possible approach to achieve this would be to
work in analogy with nonabelian bosonization and view the
extra dimensions as a (particularly large!) `internal' space
of one-dimensional fermions. Some encouraging results in this
direction were obtained but the problem is still open and
deserves further study.



\begin{thebibliography}{99}

\bibitem{Early}
F.~Bloch, Z.\ Phys.\ {\bf 81}, 363 (1933);
S.~Tomonaga, Prog.\ Theor.\ Phys.\ {\bf 5}, 544 (1950).

\bibitem{Bozo}
W.~Thirring, Ann.\ Phys.\ (N.Y.) {\bf 3}, 91 (1958);
J.~M.~Luttinger, J.\ Math.\ Phys.\ {\bf 4}, 1154 (1963);
D.\ Mattis and E.\ Lieb, J.\ Math.\ Phys.\ {\bf 6}, 304 (1965);
S.~R.~Coleman, Phys.\ Rev.\ D {\bf 11}, 2088 (1975);
S.~Mandelstam, Phys.\ Rept.\  {\bf 23} (1976) 307.

\bibitem{WZW}
E.~Witten,
Commun.\ Math.\ Phys.\  {\bf 92}, 455 (1984).

\bibitem{droplet}
J.~Polchinski, Nucl.\ Phys.\ B {\bf 362}, 125 (1991);
S.~Iso, D.~Karabali and B.~Sakita,
Phys.\ Lett.\ B {\bf 296}, 143 (1992) [arXiv:hep-th/9209003];
B.~Sakita, Phys.\ Lett.\ B {\bf 387}, 118 (1996) [arXiv:hep-th/9607047];
D.~Karabali and V.~P.~Nair, Nucl.\ Phys.\ B
{\bf 679}, 427 (2004) [arXiv:hep-th/0307281];
{\bf 697}, 513 (2004) [arXiv:hep-th/0403111].

\bibitem{phase}
A.~P.~Polychronakos,
Nucl.\ Phys.\ B {\bf 705}, 457 (2005) [arXiv:hep-th/0408194];
Nucl.\ Phys.\ B {\bf 711}, 505 (2005) [arXiv:hep-th/0411065].


\bibitem{collect}
A.~Jevicki and B.~Sakita,
Nucl.\ Phys.\ B {\bf 165}, 511 (1980);
D.~Karabali and B.~Sakita,
Int.\ J.\ Mod.\ Phys.\ A {\bf 6}, 5079 (1991);
S.~R.~Das, A.~Dhar, G.~Mandal and S.~R.~Wadia,
Mod.\ Phys.\ Lett.\ A {\bf 7}, 71 (1992) [arXiv:hep-th/9111021];
Int.\ J.\ Mod.\ Phys.\ A {\bf 8}, 325 (1993) [arXiv:hep-th/9204028].

\bibitem{Dhar}
A.~Dhar, G.~Mandal and N.~V.~Suryanarayana,
JHEP {\bf 0601}, 118 (2006) [arXiv:hep-th/0509164];
A.~Dhar and G.~Mandal, Phys.\ Rev.\ D {\bf 74}, 105006 (2006) [arXiv:hep-th/0603154].

\bibitem{ncboz}
A.P.~Polychronakos,
Phys.\ Rev.\ Lett. {\bf 96}, 186401 (2006) [arXiv:hep-th/0502150].

\bibitem{hiboz}
A.~Luther, Phys.\ Rev.\ B  {\bf 19}, 320 (1979);
F.~D.~M.~Haldane, Varenna 1992 Lectures and cond-mat/0505529;
A.~Houghton and J.~B.~Marston, Phys.\ Rev.\ B {\bf 48}, 7790 (1993) [arXiv:cond-mat/9210007];
A.~H.~Castro Neto and E.~Fradkin, Phys.\ Rev.\ Lett.\ {\bf 72}, 1393 (1994) [arXiv:cond-mat/9304014]; Phys.\ Rev.\ B {\bf 49}, 10877 (1994) [arXiv:cond-mat/9307005];
D.~Schmeltzer and A.R.~Bishop, Phys.\ Rev.\ B {\bf 50}, 12733 (1994);
D.~V.~Khveshchenko, Phys.\ Rev.\ B {\bf 52}, 4833 (1995)  [arXiv:cond-mat/9409118];
D.~Schmeltzer, Phys.\ Rev.\ {\bf 54}, 10269 (1996); D. Karabali, Nucl.\ Phys.\ B {\bf 750},  265 (2006) [arXiv:hep-th/0605006].


\bibitem{alexios1}
A. Enciso and A.P. Polychronakos,  Nucl.\ Phys.\ B {\bf 751}, 376 (2006) [arXiv:hep-th/0605040].

\bibitem{sina}
D.~B.~Fairlie and C.~K.~Zachos,
Phys.\ Lett.\ B {\bf 224}, 101 (1989).




\end{thebibliography}
\end{document}